\documentclass[prl,superscriptaddress,twocolumn,showpacs]{revtex4}
\usepackage{graphicx,dcolumn,bm,amsmath,amssymb,color,epstopdf}

\begin{document}

\title{Elucidation of the anomalous A = 9 isospin quartet behaviour}

\author{M.~Brodeur}
\email[Corresponding author:~]{brodeur@nscl.msu.edu}
\affiliation{TRIUMF, 4004 Wesbrook Mall, Vancouver BC, Canada V6T 2A3}
\affiliation{Department of Physics and Astronomy,
University of British Columbia, Vancouver, BC, Canada V6T 1Z1}
\affiliation{National Superconducting Cyclotron Laboratory, Michigan State University, East Lansing, Michigan 48824, USA}
\author{T.~Brunner}
\affiliation{TRIUMF, 4004 Wesbrook Mall, Vancouver BC, Canada V6T 2A3}
\affiliation{Physik Department E12, Technische Universit\"{a}t
M\"{u}nchen, James Franck Str., Garching, Germany}
\author{S.~Ettenauer}
\affiliation{TRIUMF, 4004 Wesbrook Mall, Vancouver BC, Canada V6T 2A3}
\affiliation{Department of Physics and Astronomy,
University of British Columbia, Vancouver, BC, Canada V6T 1Z1}
\author{A.~Lapierre}
\affiliation{TRIUMF, 4004 Wesbrook Mall, Vancouver BC, Canada V6T 2A3}
\affiliation{National Superconducting Cyclotron Laboratory, Michigan State University, East Lansing, Michigan 48824, USA}
\author{R.~Ringle}
\affiliation{TRIUMF, 4004 Wesbrook Mall, Vancouver BC, Canada V6T 2A3}
\affiliation{National Superconducting Cyclotron Laboratory, Michigan State University, East Lansing, Michigan 48824, USA}
\author{B.A.~Brown}
\affiliation{National Superconducting Cyclotron Laboratory, Michigan State University, East Lansing, Michigan 48824, USA}
\author{D.~Lunney}
\affiliation{CSNSM-IN2P3-CNRS, Universit\'{e} Paris 11, 91405
Orsay, France}
\author{J.~Dilling}
\affiliation{TRIUMF, 4004 Wesbrook Mall, Vancouver BC, Canada V6T 2A3}
\affiliation{Department of Physics and Astronomy,
University of British Columbia, Vancouver, BC, Canada V6T 1Z1}

\begin{abstract}
Recent high-precision mass measurements of  $^{9}$Li and $^{9}$Be, performed with the TITAN Penning trap at the TRIUMF ISAC facility, are analyzed in light of state-of-the-art shell model calculations.  We find an explanation for the
anomalous Isobaric Mass Multiplet Equation (IMME) behaviour for the two $A$ = 9 quartets. The presence of a cubic $d$ = 6.3(17) keV term for the $J^{\pi}$ = 3/2$^{-}$ quartet and the vanishing cubic term for the excited $J^{\pi}$ = 1/2$^{-}$ multiplet depend upon the presence of a nearby $T$ = 1/2 state in $^{9}$B and $^{9}$Be that induces isospin mixing. This is contrary to previous hypotheses involving purely Coulomb and charge-dependent effects. $T$ = 1/2 states have been observed near the calculated energy, above the $T$ = 3/2 state. However an experimental confirmation of their $J^{\pi}$ is needed.
 \end{abstract}


\maketitle

Atomic nuclei are described by their binding energy and three quantum numbers: the total angular momentum $J$, parity $\pi$, and isospin $T$. This framework allows one to identify, each of the $\sim$ 3000 observed nuclei \cite{ENS12} unambiguously. The isospin quantity is analogous to spin and was first introduced by Heisenberg \cite{Hei32} to describe the charge-independence of the nuclear force. Within the isospin formalism, neutrons (n) and protons (p) are nucleons of isospin $T$ = 1/2 but distinguished by different z-projections $T_{z}$(n) = 1/2 and $T_{z}$(p) = -1/2 \cite{Hei32, Wig37}. Nuclei with the same mass number $A$, total angular momentum and parity form multiplets where the individual members have a projection \(T_{z} = (N-Z)/2\). Assuming isospin is a good quantum number, members of an isobaric multiplet have identical properties. However, Weinberg and Treiman \cite{Wei59} noted that the mass excess $\Delta$ (which is a measure of the nuclear binding energy and defined as the difference between the atomic mass and the atomic mass number) of such nuclides were not identical, but were rather laying along a parabola:
\begin{equation}\label{eq:IMME1}
\Delta(A,T,T_{z}) = a(A,T) + b(A,T)T_{z} + c(A,T)T_{z}^{2}
\end{equation}  
where $a$, $b$, $c$ are coefficients that depend on all quantum numbers except $T_z$. This so-called isobaric multiplet mass equation (IMME) has proven to be a powerful tool to predict unknown masses. For instance, it is used to obtain masses of nuclei along the rapid proton capture path, where most of the masses are not well known \cite{Sch06} or to provide detailed mass values, which are experimentally inaccessible due to half-life and productions constraints \cite{Ade99}. Recently, the precise mass measurement of $^{12}$Be \cite{Ett10} using the TITAN (TRIUMF Ion Traps for Atomic and Nuclear science) Penning trap mass spectrometer \cite{Dil03,Dil06} has been used as a solid anchor point together with the IMME to address the ambiguous spin assignment of $T$ = 2 states in $^{12}$C and $^{12}$Be.

Several tests of the IMME were performed and for most cases, it has followed the original quadratic behaviour \cite{Bri98}. However, in some cases, large deviations were discovered and the incorporation of cubic $d(A,T)T_{z}^{3}$ and/or quartic $e(A,T)T_{z}^{4}$ \cite{Jan69, Ber70} terms was considered. The largest documented breakdowns include the $A$ = 9 \cite{Kas75} and $A$ = 33 \cite{Her01} quartets and the $A$ = 8 \cite{Cha11} and $A$ = 32 \cite{Kwi09,Sig11} quintets. The unveiling of the non-quadratic behaviour of the $A$ = 32 and 33 multiplets was only possible due to the precise and accurate mass measurement of some of its members, at the $\delta m/m \sim$ 10$^{-7}$ level, using Penning traps \cite{Bla06}. Because of their lighter mass, the $A$ = 8 and 9 multiplet breakdowns where discovered sooner, from less precise reaction $Q$-value mass determinations \cite{Kou75,Kas74}. More recently, mass measurements of $^{8}$C \cite{Cha11b} and $^{8}$He \cite{Bro12} showed the need for a larger cubic $d$ = 11.1(2.3) keV term in the $A$ = 8 quintet \cite{Cha11}. In addition, a new IMME evaluation of the $A$ = 9 quartet became possible using recent lithium \cite{Smi08} and beryllium \cite{Rin09} mass measurements. The $A$ = 9 isobars are of particular interest because it is the first and lightest chain presenting two different IMME quartets \cite{Kas74,Ben74}. The ground state quartet \cite{Kas74} strongly departs from quadrature with a quadratic fit $\chi^{2}$ of 10.2 and a cubic coefficient $d$ = 5.5(18) keV \cite{Bri98}. Several mechanisms to explain this departure have been proposed including the Coulomb-dependent Thomas-Ehrman shift \cite{Lan58} arising from the small binding energy of the last proton in $^{9}$C and non-Coulomb charge-dependant forces \cite{Ber70}. However, to date, the total contribution from these mechanisms was insufficient to explain the observed cubic term \cite{Ber70,Kas75}. Furthermore, as the strength of these effects would increase with a decreasing proton separation energy in $^{9}$C, the excited state quartet should depart more strongly from quadrature \cite{Ber70}. However, the excited state quartet in $A$ = 9 shows a good agreement with quadrature ($\chi^{2}$ = 1.0) and a nearly vanishing cubic term $d$ = 3.5(34) keV \cite{Bri98}. The mixing of the $T$ = 3/2 and $T$ = 1/2 states for the $T_z$ = $\pm$1/2 members has been proposed \cite{Jan69}, but the large width of the known resonances met with skepticism \cite{Har71} for the strength of this mechanism. This Letter explains the long-standing enigmatic behaviour of the two $A$ = 9 quartets.
\begin{table}[ht]
\begin{center}
\caption{\label{tab:NewMass} New ground state mass excesses $\Delta_{\rm G.S.}$ and excitation energies \cite{ENS12} of the $T$ = 3/2 states used for the calculation of the $J^{\pi}$ = 3/2$^{-}$ and 1/2$^{-}$ $A$ = 9 quartets. The $^{9}$B $J^\pi$ = 1/2$^{-}$ excitation energy labeled with (*) is from \cite{Cha11}.}
\begin{tabular}{lcllll}
\hline
& $T_z$ & $\Delta_{\rm G.S.}$ & $E_{x}$(3/2$^{-}$) & $E_{x}$(1/2$^{-}$)  \\
& & (keV) & (keV) & (keV) \\
\hline 
$^{9}$Li & 3/2 & 24 954.91(20) & 0 & 2 691(5)  \\
$^{9}$Be & 1/2 & 11 348.391(93) & 14 392.2(18) & 16 977.1(5) \\
$^{9}$B & -1/2 & 12 416.4(10) & 14 655.0(25) & 17 076(4) \\
 & & & & 16 990(30) * \\
$^{9}$C & -3/2 & 28 909.5(21) & 0 & 2 218(11) \\
\hline 
\end{tabular}
\end{center}
\end{table}
\begin{figure}[t]
\begin{center}
\includegraphics[width=0.5\textwidth,clip=]{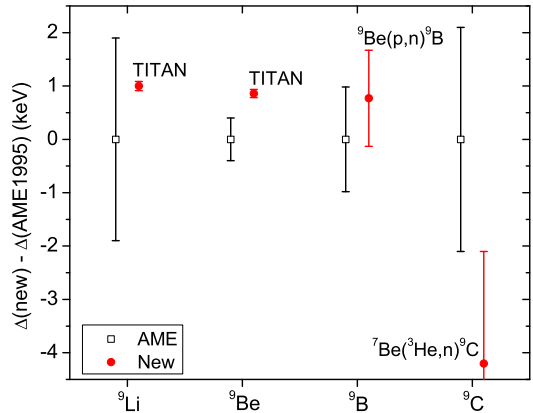}
\caption{(Color on-line) New mass excess $\Delta$(new) values presented in table~\ref{tab:NewMass} compared to the 1995 atomic mass evaluation (AME1995) \cite{Aud95} mass excess values $\Delta$(AME1995) used in the most recent IMME review \cite{Bri98}. The new $^{9}$Li and $^{9}$Be $\Delta$ were measured at TITAN, while the $^{9}$B and $^{9}$C $\Delta$ involve respectively a $^{9}$Be(p,n)$^{9}$B and $^{7}$Be($^{3}$He,n)$^{9}$C reaction $Q$-value. \label{fig:AMEdiff}}
\end{center}
\end{figure}

The most recent IMME review (1998), used the ground state masses from the 1995 atomic mass evaluation (AME1995) \cite{Aud95}. In this evaluation, all ground state masses for the $A$ = 9 quartets where based on reaction $Q$-value measurements from transfer reactions. Since then, the masses of $^{9}$Li  \cite{Smi08} and $^{9}$Be \cite{Rin09} where measured directly using the TITAN Penning trap mass spectrometer of the ISAC facility at TRIUMF. Penning traps have been established as the most precise and reliable devices for mass measurements \cite{Bla06}, which is a key prerequisite for this IMME study. The TITAN system has been established with mass measurements of halo nuclei \cite{Ryj08, Smi08, Rin09, Bro12} as well as with measurements of short-lived, highly-charged ions \cite{Ett11}. The precision and accuracy of the TITAN Penning trap has been studied in detail \cite{Bro09,Bro12b}. 

The new masses are presented in table~\ref{tab:NewMass}. The large improvement in precision and change in the mass value compared to AME1995 is displayed in Fig.~\ref{fig:AMEdiff}. The TITAN $^{9}$Be mass excess also affects $^{9}$B, which is derived from the $Q$-value of a $^{9}$Be(p,n)$^{9}$B reaction \cite{Aud93, Aud03}. The 4.2 keV decrease in the $^{9}$C mass excess, currently derived from a $^{7}$Be($^{3}$He,n)$^{9}$C $Q$-value, originates from the change in the $^{7}$Be mass evaluation \cite{Aud93, Aud03}. The $^{7}$Be mass value in the 2003 Atomic Mass Evaluation (AME03) was based solely on the $Q$-value of the reaction $^{7}$Li(p,n)$^{7}$Be \cite{Aud03} which was not the case for the 1995 evaluation \cite{Aud93}. The $^{9}$C mass presented in table~\ref{tab:NewMass} is also based on this reaction, but uses the precise $^{7}$Li mass value measured using the SMILETRAP Penning trap spectrometer \cite{Nag06}. All the excitation energies presented in table~\ref{tab:NewMass} come from the Evaluated Nuclear Structure Data File (ENSDF) \cite{ENS12}, except for the $^{9}$B $J^\pi$ = 1/2$^{-}$ state. For this state we include a recent 3$\sigma$ lower value of 16 990(30) keV \cite{Cha11b} together with the 17 076(4) keV value from ENSDF. In the new measurement, the angular correlation between the first proton emission of $^{9}$B and subsequent double $\alpha$ decay strongly suggest the $J^\pi$ = 1/2$^{-}$ assignment for this state. 

\begin{table}[ht]
\begin{center}
\caption[IMME coefficients using the TITAN masses.] 
{\label{tab:IMMEfit} IMME coefficients (equation~\eqref{eq:IMME1}) using the mass excesses calculated from the values in table~\ref{tab:NewMass}. Also given are the $\chi^{2}$ and P-value of the various fits. The (*) value used the $^9$B $J^\pi$ = 1/2$^{-}$ excitation energy from \cite{Cha11b}.}
\begin{tabular}{lclllll}
\hline
$J^\pi$ & $a$ & $b$ & $c$ & $\chi^{2}$ & P-values \\
& (keV) & (keV) & (keV) & &  \\ 
\hline 
3/2$^{-}$ & 26337.5(17) & -1318.8(7) & 264.7(9) & 14.6 & 1$\times$10$^{-4}$ \\
1/2$^{-}$ & 28847.1(15) & -1163.7(29) & 241.2(24) & 1.2  & 3$\times$10$^{-1}$ \\
1/2$^{-}$ * & 28845.0(19) & -1159.1(41) & 240.1(25) & 6.9 & 9$\times$10$^{-3}$ \\
\hline
\end{tabular}
\end{center}
\end{table}

The updated IMME coefficients for the fits, including quadratic and cubic terms are shown in Table~\ref{tab:IMMEfit} together with their $\chi^{2}$. As the various fits involve only one degree of freedom, we also give the corresponding P-value for the $\chi^{2}$. The new mass excesses result in a 40$\%$ increase in the $\chi^{2}$ for the $J^\pi$ = 3/2$^{-}$ quartet quadratic fit compared to the previous value \cite{Bri98}. This very large $\chi^{2}$ is associated with a small P-value, which reflects the probability to obtain such a $\chi^{2}$ by accident. Performing a cubic fit results in an enhanced $d$ coefficient of 6.3(17) keV. The excited $J^\pi$ = 1/2$^{-}$ quartet, on the other hand, shows a good agreement with quadrature leading to a $\chi^{2}$ = 1.2 when using the ENSDF $^{9}$B excitation energy \cite{ENS12}. However, if the 16 990(30) keV value for the $^{9}$B $J^\pi$ = 1/2$^{-}$ excitation energy from \cite{Cha11b} is used, the excited state quartet also departs from quadrature with a $\chi^{2}$ = 6.9 and a larger cubic term of $d$ = -40(15) keV. A more precise confirmation of this recent measurement is desired, to resolve this discrepancy for the excited quartet.

\begin{table}
\begin{center}
\caption{\label{tab:b9} Experimental and theoretical $d$, $b_1$ and $b_3$ coefficients for the two $A$ = 9.}
\begin{tabular}{|c|lll|lll|}
\hline
        & $  b_{1}(3/2^{-})  $  & $  b_{3}(3/2^{-})  $ & $  d(3/2^{-})  $
        & $  b_{1}(1/2^{-})  $  & $  b_{3}(1/2^{-})  $ & $  d(1/2^{-})  $ \\
        & (keV) & (keV) & (keV) & (keV) & (keV) & (keV) \\
\hline
exp & -1330.8(32) & -1318.2(7) & 6.3(17)   & -1161(4) & -1167(4) & 3.2(2.9) \\
PJT & -1334  & -1321  & 6(2)       & -1258  & -1260 &  -1(2)  \\
CKI & -1364  & -1342  & 11(2)      & -1290   & -1290  &   0(2) \\
\hline
\end{tabular}
\end{center}
\end{table}

In order to see how the large cubic term of the ground state quartet arises, we define two new $b$ coefficients, $b_{1}$ and $b_{3}$:
\begin{eqnarray}\label{eq:IMME2}
b_{1} = \Delta (A,T_{z}=1/2) - \Delta (A,T_{z}=-1/2) \\
b_{3} = [\Delta (A,T_{z}=3/2) - \Delta (A,T_{z}=-3/2)]/3. 
\end{eqnarray}  
The $b$ and $d$ coefficients can be written in terms of $b_{1}$ and $b_{3}$ as \(d = [b_{1} - b_{3}]/2\) and \(b = [9 b_{1} - b_{3}]/8\). From these expressions, it can be seen that for \(b_{3} = b_{1}\), the other coefficients become $d$ = 0 and \(b = b_{3} = b_{1}\). With this change, the physics involving only the $T_z$ = $\pm$1/2 states is now explicitly contained in the single $b_1$ term. The experimental and calculated results for these coefficients are presented in table~\ref{tab:b9}. The results were calculated using a shell model Hamiltonian comprising two parts: isospin conserving and isospin non-conserving \cite{ormand} component. The first part uses either the Cohen-Kurath two-body matrix element (6-16)2BME (CKI) \cite{cki} or the updated version of this, the so-called p-shell potential model fit (PJT) \cite{pjt}. The PJT Hamiltonian is given in \cite{pjtb}. Both of these are based on fits of the two isospin-conserving single-particle energies and fifteen two-body matrix elements adjusted to binding energies and excitation energies for the $p$-shell nuclei from $A$ = 6 to 16. The PJT Hamiltonian takes into account more data than was available for the CKI. The isospin non-conserving Hamiltonian includes a Coulomb, charge-asymmetric isovector and charge-dependent isotensor strong interaction. The free parameters of these interactions were obtained by fitting the Hamiltonian to the $p$-shell nuclei experimental data for the $b$ and $c$ coefficients of the IMME.
\begin{figure}
\begin{center}
\includegraphics[width=0.5\textwidth,clip=]{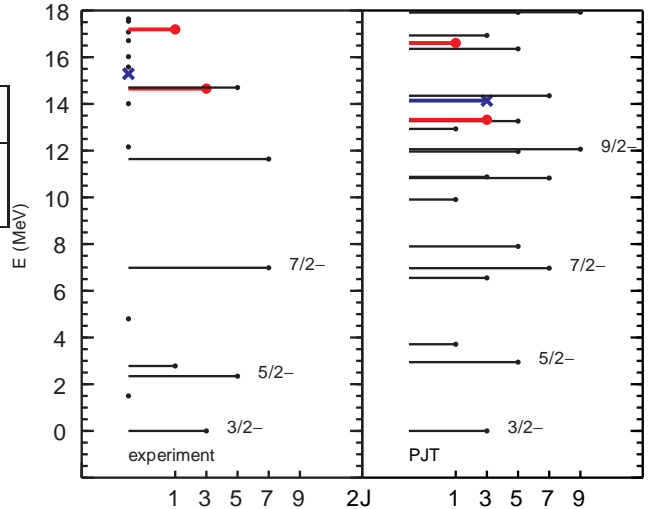}
\caption{(Color on-line) Experimental and theoretical energies for
negative parity levels of $^{9}$B.
The total angular momenta $J$ are indicated by the length of the line.
The experimental levels for which $J$ is uncertain are indicated
by the black dots. The states indicated by a thick red line ending with a circle are the $T$ = 3/2  
states of interest. The theoretical state indicated by a thick blue line ending with a cross is the $T$ = 1/2  
state that isospin mixes with the lower $T$ = 3/2 state. The
experimental state in blue is a possible candidate for the
associated $T$ = 1/2, $J^{\pi}$ =  3/2$^{-}$ level.\label{fig:b9} 
}
\end{center}
\end{figure}
The calculated energy levels, total angular momentum and parity using the PJT
Hamiltonian are compared to the experimental values in Fig.~\ref{fig:b9}. The following discussion will focus on
the $J^{\pi}$ =  1/2$^{-}$ and 3/2$^{-}$, $T$ = 3/2   states indicated by thick red lines in Fig.~\ref{fig:b9}.

Table~\ref{tab:b9} shows that the theoretical results for the $d$ coefficient, calculated to one keV precision, are in excellent agreement with experiment. The better agreement for the PJT potential could be attributed to its larger available data set for fitting. The non-zero cubic term for the $J^{\pi}$ =  3/2$^{-}$, $T$ = 3/2  level is dominated by isospin mixing with a higher $J^{\pi}$ =  3/2$^{-}$, $T$ = 1/2 state. Calculations using the CKI Hamiltonian predicts this level to be about 0.5 MeV higher, while the PJT Hamiltonian predicts this level to be about 0.8 MeV higher (right-hand side Fig.~\ref{fig:b9}). Experimentally, a $T$ = 1/2 level with unknown total angular momentum, parity and width is observed very close to the $T$ = 3/2 level expected from PJT, at 0.6 MeV (left-hand side of Fig.~\ref{fig:b9}) \cite{ENS12}. This isospin mixing can be understood as a second-order perturbation in which these two levels repel each other. As a result, the $T$ = 3/2 level is pushed down by an amount proportional to the expectation value of the isospin non-conserving interaction divided by the energy difference $\Delta E$ between the two states: \(\mid <V_{{\rm INC}}>\mid ^{2}/\Delta E\). The difference in the predicted $d$ coefficient from the CKI and PJT Hamiltonian is mainly due to the different energy denominator. If the mixing originates from the observed $T$ = 1/2 state suggested above, then the PJT results would be in good agreement with experiment for both $\Delta E$ and $d$. In contrast, calculations using either Hamiltonian do not predict a $J^{\pi}$ =  1/2$^{-}$, $T$ = 1/2 state near the quartet $J^{\pi}$ =  1/2$^{-}$, $T$ = 3/2 state, leading to a nearly zero $d$ coefficient. This is confirmed experimentally as no nearby $T$ = 1/2 states have been observed \cite{ENS12}.

The interference between the Coulomb, charge-asymmetric isovector and charge-dependent isotensor part of the isospin non-conserving interaction is the critical ingredient that gives rise to the level repulsion stemming from the isospin mixing. If the Coulomb term is put to zero and either the isotensor or isovector part of the interactions is used, then the corresponding matrix element is the same for both $^{9}$Be and $^{9}$B, resulting in an equal downward push of their respective $T$ = 3/2 states and a $d$ = 0 term. In detail, the calculation decomposes these
matrix elements into a sum of Coulomb, charge-asymmetric isovector and 
charge-dependent isotensor contributions resulting in
$\mid <V_{{\rm INC}}>\mid$ = $\mid$ 54 $+$ 44 $+$ 18 $\mid$ = 116 keV and
$\mid <V_{{\rm INC}}>\mid$ = $\mid -$ 19 $-$ 44 $+$ 18 $\mid$ = 45 keV
for $^9$B and $^9$Be respectively. The square of these values divided by the energy difference between the $T$ = 3/2 and nearby $T$ = 1/2 state yield downward shifts of about 14 keV in $^{9}$B and 3 keV in $^{9}$Be. Then, the $d$ coefficient is equal to half the difference between the two downward shifts. 

The isospin non-conserving matrix elements theoretical uncertainties are related to the model and uncertainties used for fitting the $b$ and $c$ coefficients as given in Table 2 of \cite{ormand}. The Coulomb term is well determined and is within 5$\%$ of its expected strength. The $c$-coefficient is sensitive to the charge-dependent isotensor term \cite{Aue83}. Its value and error are given in Table 2 of \cite{ormand} as $S_{0}^{(2)}$ = -0.017(5), leading to an error of about 1 keV in the $d$-coefficient. The charge-asymmetric term was determined from a fit to the $b$-coefficients and given in Table 2 of \cite{ormand} as $S_{0}^{(1)}$ = -0.042(11), leading to an error of about 1 keV in the $d$-coefficient. Both of these contributions yield a total theoretical uncertainty of 2 keV. In \cite{ormand} these empirical coefficients were compared to simple model estimates based on the experimental nucleon-nucleon scattering lengths; $S_{0}^{2}$ = -0.042(12) and $S_{0}^{1}$ = -0.006(4). Using these estimates in place of the empirical values results in $d$ = 7 keV; essentially the same result although the details in terms of the division between the isovector and isotensor terms is rather different. In the future it would be desired that the isospin non-conserving contributions used for the $\mid <V_{{\rm INC}}>\mid$ should be reevaluated by renormalizing nucleon-nucleon interactions to the $p$-shell model space with inclusion of three-body interactions.

Previous work on the theoretical calculations of the IMME including the $d$ coefficients is reviewed in \cite{Aue83}. Calculations of the $A$ = 9 $J^+$ = 3/2$^-$ quartet was considered in \cite{Ber70} where the Coulomb interaction was used in a schematic model, yielding $d$ = 0.9 keV. Charge-dependent interactions resulted in an extra 1 keV. Most importantly, the results obtained in \cite{Ber70} do not include the contribution from mixing with specific nearby states and they are thus not inconsistent with our result as our larger cubic term is primarily due to the mixing with a nearby state.

In summary, combining high-precision $^9$Li and $^9$Be mass measurements performed using the TITAN Penning trap spectrometer, and shell model calculations using the most complete potential to date, we could fully explain for the first time the binding energy behaviour of the $A$ = 9 quartet. Contrary to previous hypothesis, higher-order charge dependant effects and Coulomb effect-induced expansion of the wave function are not the leading mechanism driving a non-zero term in this quartet. It is created by the isospin mixing of the $T$ = 3/2 level with an above $T$ = 1/2 level. Using this approach, the long-standing anomalous difference in $d$-coefficients for the two $A$ = 9 quartet is explained by the absence of a nearby $T$ = 1/2 level for the $J^{\pi}$ = 1/2$^-$ quartet. Nevertheless, an experimental confirmation of our predicted $J^{\pi}$ for the 15 100(50) keV and 15 290(40) keV states in $^9$Be and $^9$B  as well as a measurement of their width would be of interest. Also, light should be shed on the excitation energy discrepancy for the $^9$B $J$ = 1/2$^-$ level. 

Finally, the presented interpretation for the creation of a non-zero $d$ coefficient in the ground-state $A$ = 9 quartet can be generalized to other isobar multiplets if the same conditions of a small energy separation between mixing levels and a non-negligible isospin non-conserving matrix element is met. This generalized theoretical description can also be used to explain other physical effects such as isospin-forbidden proton decays \cite{Orm86} and the necessity of a correction term $\delta_c$ to properly determine the Cabibbo-Kobayashi-Maskawa matrix $V_{ud}$ mixing angle \cite{Tow10, Sig11}.  

This work was supported by the Natural Sciences and Engineering
Research Council of Canada (NSERC) and the National Research Council
of Canada (NRC). We would like to thank the TRIUMF technical staff,
especially Melvin Good. S.E.~acknowledges support from the Vanier
CGS program, T.B.~from the Evangelisches Studienwerk e.V.~Villigst and
D.L.~from TRIUMF.

\bibliography{apssamp}

\end{document}